\documentclass[sigconf]{acmart}

\definecolor{brown}{rgb}{0.59, 0.29, 0.0}
\definecolor{darkgray}{rgb}{0.59, 0.59, 0.59}
\definecolor{tablegray}{gray}{.9}

\newcommand\rv[1]{#1}

\newcommand{\system}{\textsc{Sensonaut}}
\newcommand{\initangle}{\texttt{ANGLE}}
\newcommand{\numobjs}{\texttt{NUM\_OBJS}}
\newcommand{\numdist}{\texttt{DISTRACTORS}}

\usepackage[utf8]{inputenc}
\usepackage{diagbox}
\usepackage{colortbl}
\usepackage{booktabs}
\usepackage{caption}

\usepackage{color}
\usepackage{soul}
\usepackage{lipsum}
\setlipsum{%
  par-before = \begingroup\color{gray},
  par-after = \endgroup
}
\usepackage{tabularx}

\usepackage{nameref}

\usepackage{xspace}
\usepackage{enumitem}
\usepackage{mathtools}
\usepackage{commath}
\usepackage{amsmath}

\usepackage{enumitem}
\usepackage{placeins}

\usepackage{amssymb}
\usepackage{pifont}

\newcommand{\customtilde}{{\raise.17ex\hbox{$\scriptstyle\sim$}}}

\newcommand{\eg}{e.\,g.,\xspace}
\newcommand{\ie}{i.\,e.,\xspace}

\usepackage{xparse}
\usepackage{gensymb}

\sethlcolor{yellow}

\AtBeginDocument{%
  }

\copyrightyear{2026}
\acmYear{2026}
\setcopyright{cc}
\setcctype{by}
\acmConference[CHI '26]{Proceedings of the 2026 CHI Conference on Human Factors in Computing Systems}{April 13--17, 2026}{Barcelona, Spain}
\acmBooktitle{Proceedings of the 2026 CHI Conference on Human Factors in Computing Systems (CHI '26), April 13--17, 2026, Barcelona, Spain}
\acmPrice{}
\acmDOI{10.1145/3772318.3790614}
\acmISBN{979-8-4007-2278-3/2026/04}

\begin{document}

\title{Simulating Human Audiovisual Search Behavior}

\author{Hyunsung Cho}
\orcid{0000-0002-4521-2766}
\affiliation{%
  \institution{Aalto University}
  \city{Helsinki}
  \country{Finland}}
\affiliation{%
  \institution{Carnegie Mellon University}
  \city{Pittsburgh}
  \state{PA}
  \country{USA}}
\email{hyunsung@cs.cmu.edu}

\author{Xuejing Luo}
\orcid{0009-0004-7520-980X}
\affiliation{%
  \institution{Aalto University \& ELLIS Institute}
  \city{Helsinki}
  \country{Finland}}
\email{xuejing.luo@aalto.fi}

\author{Byungjoo Lee}
\orcid{0009-0003-1547-9923}
\affiliation{
    \institution{Yonsei University}
    \city{Seoul}
    \country{Republic of Korea}}
\email{byungjoo.lee@yonsei.ac.kr}

\author{David Lindlbauer}
\orcid{0000-0002-0809-9696}
\affiliation{
  \institution{Carnegie Mellon University}
  \city{Pittsburgh}
  \state{PA}
  \country{USA}}
\email{davidlindlbauer@cmu.edu}

\author{Antti Oulasvirta}
\orcid{0000-0002-2498-7837}
\affiliation{
  \institution{Aalto University \& ELLIS Institute}
  \city{Helsinki}
  \country{Finland}}
\email{antti.oulasvirta@aalto.fi}

\renewcommand{\shortauthors}{Cho et al.}

\begin{abstract}
Locating a target based on auditory and visual cues---such as finding a car in a crowded parking lot or identifying a speaker in a virtual meeting---requires balancing effort, time, and accuracy under uncertainty.
Existing models of audiovisual search often treat perception and action in isolation, overlooking how people adaptively coordinate movement and sensory strategies.
We present \system{}, a computational model of \textit{embodied audiovisual search}.
The core assumption is that people deploy their body and sensory systems in ways they believe will most efficiently improve their chances of locating a target, trading off time and effort under perceptual constraints.
Our model formulates this as a resource-rational decision-making problem under partial observability.
We validate the model against newly collected human data, showing that it reproduces both adaptive scaling of search time and effort under task complexity, occlusion, and distraction, and characteristic human errors.
Our simulation of human-like resource-rational search informs the design of audiovisual interfaces that minimize search cost and cognitive load.
\end{abstract}

\ccsdesc[500]{Human-centered computing~HCI theory, concepts and models}

\keywords{Computational behavior modeling; user simulation; multimodal perception; computational rationality; reinforcement learning}

\begin{teaserfigure}
  \includegraphics[width=\textwidth]{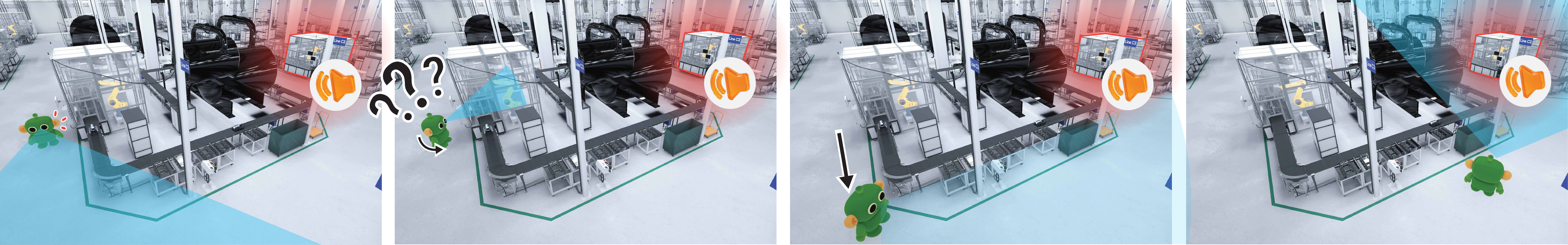}
  \caption{
  \system{} is a model for embodied audiovisual search behavior. It simulates how humans resolve perceptual ambiguities such as noise, occlusion, and confusion through embodied actions of head rotation and locomotion. This model can be used to predict users' physical effort and time in locating a spatial sound source in various HCI applications.}
  \Description{}
  \label{fig:teaser}
\end{teaserfigure}

\maketitle
\section{Introduction}
Supporting people to locate objects in space is a long-standing challenge for HCI research~\cite{finnegan2016compensating,evangelista2022auit}.
In ubiquitous computing, this includes helping a driver find their car in a crowded parking lot, navigating a visitor to a meeting room, or guiding a user in high-stakes settings, such as a nurse in a busy intensive care unit~(ICU) or a first responder in an industrial facility~(\autoref{fig:teaser}).
In immersive environments, it extends to digital elements embedded in the surrounding scene, such as floating widgets, notifications, or avatars.
Although simple directional guidance may suffice under clear conditions, guidance cues can often appear ambiguous: sounds can be masked by background noise, objects can be hidden behind others, and multiple items can look or sound similar.
To resolve this uncertainty, people do not passively wait for instructions; they actively move their eyes, heads, and bodies to gather information over time. 
Consequently, effective spatial assistance cannot rely on static cues alone but must adapt to the user’s unfolding behavior and physical constraints.
We therefore argue that a computational model of this behavior is essential for predicting when assistance will succeed or fail, and how it might be improved. 

However, current computational models focus either on low-level perceptual localization or on abstract cognitive architectures that do not capture embodied strategies. 
For example, auditory localization models explain how people estimate the direction of sound from binaural cues~\cite{middlebrooks1991s,blauert1997spatial}, but do not extend to predicting how users move and search sequentially to estimate.
An ideal observer model based on Bayesian inference frameworks such as recursive Bayesian estimation~\cite{mclachlan2025bayesian} captures how sensory evidence accumulates over time but typically abstracts away the embodied actions and resource constraints that shape real-world search.
What is missing is a framework that unifies sensory inference with resource-rational decision-making, capturing how people integrate uncertain cues and embodied actions when searching for targets under partial observability.

To close this gap, we present \system{}, a novel computational model of embodied audiovisual search that simulates how people locate targets under uncertainty.
We formalize this task as a partially observable Markov decision process (POMDP), where users must make sequential decisions about where to look and move based on incomplete sensory information about target locations.
The model takes multisensory observations, including auditory cues from interaural time differences~(ITD) and visual glimpses, and integrates multisensory observations with prior to update a belief distribution \rv{over potential locations}.
Using reinforcement learning, we train the model to approximate a \emph{resource-rational} policy: one that balances the value of information against the physical cost of action~(\eg turning the head or walking).
By simulating thi sprocess, the model produces emergent search trajectories, predicting not only success rates but also the time and effort to find a target and the likelihood of specific errors.

We validated the model against newly collected behavioral data from a controlled Virtual Reality (VR) study. 
In this study, participants searched for sound-emitting targets under varying number of occluding objects, number of distractor objects, and initial target direction. 
The model reproduced key performance trends, including the impact of distractors and target location on search difficulty.
Crucially, the model captured the embodied nature of the task, mirroring human strategies by prioritizing low-cost head rotations to gather information and using locomotion to resolve ambiguity only when necessary
It also replicates characteristic human modes of search error.
These results demonstrate that our model provides a resource-rational account of audiovisual search, offering a simulation tool for designing audiovisual interfaces. 
Our implementation and dataset are available as open source at \url{https://augmented-perception.org/publications/2026-sensonaut.html}.

In summary, our contribution is three-fold:
\begin{itemize}
    \item A resource-rational model of embodied audiovisual search that unifies cue integration with decision-making under embodied action costs.
    \item An implementation of this model as a POMDP solved with reinforcement learning, capable of simulating human-like search strategies and errors.
    \item A human behavioral dataset of audiovisual search tasks with systematically varied task complexity. 
\end{itemize}

\section{Background and Related Work}

\subsection{Human Audiovisual Localization and\\Cue Integration}
Humans localize sound sources by integrating multiple auditory cues, most prominently interaural time differences~(ITD) and interaural level differences~(ILD), along with spectral filtering introduced by the head and pinnae~\cite{middlebrooks1991s,blauert1997spatial}.
Duplex theory explains how ITD dominates at low frequencies while ILD dominates at high frequencies~\cite{middlebrooks1991s}, but in broadband conditions ITD often provides the most reliable azimuthal information~\cite{shinn2000tori,May_van_de_Par_Kohlrausch_2011}.
Humans reduce ambiguities such as front-back confusions by actively monitoring their heads and bodies, leveraging embodied actions to transform the acoustic input~\cite{Wallach_1940,brimijoin2012role}.

Localization rarely relies on audition alone: visual cues constrain auditory estimates and reduce uncertainty.
The classic ventriloquist effect shows that people perceptually fuse auditory and visual signals, with vision often dominating when the two conflict~\cite{Alais_Burr_2004} and consistent with Bayesian cue combination models~\cite{Ernst_Banks_2002}.
This also introduces challenges in audiovisual search: if a visual distractor resembles the target, or if another sound coincides near a distractor’s location, observers may misattribute the auditory source to the wrong visual object. 
Such crossmodal binding can therefore both sharpen localization when cues align and create systematic confusions when they do not.
Computational accounts explain this as near-optimal reliability-weighted fusion of multisensory evidence~\cite{battaglia2003bayesian,fetsch2012neural,angelaki2009multisensory}.
Cue integration is thus not passive but shaped by embodied strategies: reorientation and locomotion bring ambiguous sources into view, reducing uncertainty.
These findings emphasize that audiovisual search is inherently active, emerging from the interplay of multisensory fusion and embodied exploration~\cite{mcanally2014sound,mclachlan2025bayesian}.

\begin{figure*}[t]
    \centering
    \includegraphics[width=0.9\linewidth]{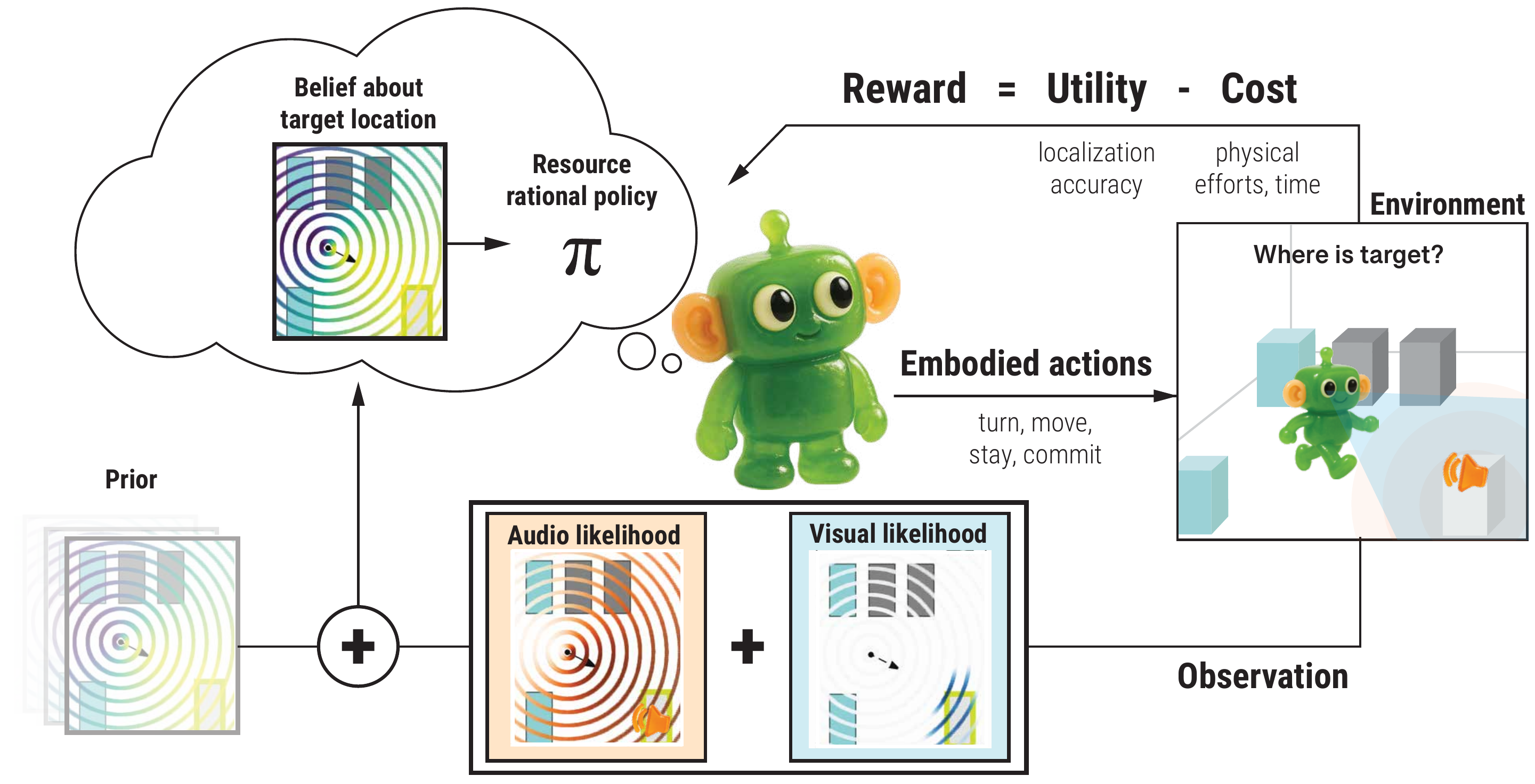}
    \caption{Overview of our computational model of embodied audiovisual search. The agent maintains a belief over possible target locations by integrating prior knowledge with audio and visual likelihoods. This belief is updated step by step and passed to a resource-rational policy, which weighs expected utility against the costs of physical effort and time. The policy selects embodied actions (turns, forward steps, staying, or committing) that shape new observations from the environment. Through this loop, the model simulates how people combine cue integration with action costs to efficiently locate targets under uncertainty.}
    \Description{}
    \label{fig:model-architecture}
    \Description{Shows an overview diagram of our computational model of embodied audiovisual search. The agent, in the middle, maintains a belief over possible target locations by integrating prior knowledge with audio and visual likelihoods. This belief is updated step by step and passed to a resource-rational policy, which weighs expected utility against the costs of physical effort and time. The policy selects embodied actions (turns, forward steps, staying, or committing) that shape new observations from the environment. Through this loop, the model simulates how people combine cue integration with action costs to efficiently locate targets under uncertainty.}
\end{figure*}




\subsection{Computational Models of Embodied Audiovisual Search}
A long tradition of computational modeling has formalized how people integrate sensory cues under uncertainty.
Bayesian ideal observer models predict how an agent should optimally combine likelihoods and priors to localize a source~\cite{Reijniers_Vanderelst_Jin_Carlile_Peremans_2014,barumerli2023bayesian}.
While insightful, these models typically assume passive observation, neglecting how human listeners actively reorient to improve information. 
Recent work extends Bayesian inference into active sensing frameworks, where actions such as head turns are selected to maximize expected information gain~\cite{najemnik2005optimal,nelson2006sensory}.

Such problems can be formalized as Partially Observable Markov Decision Processes~(POMDPs), where sequential decisions transform both observations and beliefs~\cite{drugowitsch2012cost,mclachlan2025bayesian}.
Because solving POMDPs analytically is intractable in rich environments, reinforcement learning~(RL) has been applied to approximate resource-rational strategies~\cite{binz2022modeling,Lieder_Griffiths_2020}.
RL-based agents learn when to reorient, move, or commit, balancing accuracy with the costs of time and effort~\cite{giannakopoulos2021deep,ledder2025audio,chen2020soundspaces,chen2021semantic}. 
However, \rv{most prior work addresses settings such as robotic navigation or single-modality auditory localization, rather than audiovisual search as performed by humans.
These tasks typically involve goal-directed movement toward known targets or the estimation of a single source location, often in environments with limited perceptual ambiguity.}
\rv{In contrast, human-like audiovisual search} requires fusing heterogeneous cues (\eg auditory ITDs, visual features, and occlusion structure), maintaining uncertainty over multiple potential targets and distractors, and making embodied decisions under time pressure. 
\rv{Unlike robotic navigation tasks that emphasize spatial planning or localization accuracy, our model captures human-like behaviors in embodied audiovisual search under perceptual and resource constraints.}

\subsection{Computational Models of Audiovisual Perception in HCI}
Within HCI, computational models of perception often focus on predicting noticeability and attention in multimodal interfaces~\cite{li2024predicting,cheng2025sensing,cho2025evaluating}.
Visual saliency models predict which display regions will attract gaze~\cite{wolfe2021guided} and have been extended to XR to support adaptive interfaces~\cite{halverson2011computational,chen2021adaptive}.
In parallel, auditory attention models estimate how salient sounds guide focus in cluttered environments~\cite{kaya2017modelling,bouvier2023revealing}.
Other work explores how crossmodal cues improve awareness of system events~\cite{do2020improving,cho2024sonohaptics}
Recent work such as Auptimize~\cite{cho2024auptimize} explicitly optimizes the placement of spatial audio cues in XR to minimize perceptual confusion and maximize discriminability, highlighting how computational accounts of multisensory perception can guide the design of interactive systems. 
However, these approaches typically focus on cue placement or signal design rather than the \textit{embodied dynamics} of search.  
Our contribution builds on this foundation by extending the scope from cue optimization to full embodied search: rather than focusing on where a cue should be placed, we model how users dynamically combine auditory and visual signals, update beliefs, and take embodied actions to resolve uncertainty. 
This broader account allows us to explain not only perceptual sensitivities but also the strategic, resource-rational behaviors that emerge in immersive search tasks.
\section{Model of Embodied Audiovisual Search}

In this section, we introduce \system{}, a computational model of embodied audiovisual search~(\autoref{fig:model-architecture}).
The model simulates how humans integrate auditory and visual cues and take embodied actions~(\eg turning head or moving forward) to reduce uncertainty during search. 
We instantiate this process through an \textit{agent} that perceives the environment through noisy auditory and visual cues and chooses embodied actions under uncertainty. 
This simulates how humans balance the value of committing to a correct location against the costs of turning, moving, or waiting, which can be modeled through a resource-rational account of audiovisual search.
This allows us to explain and predict how users behave in audiovisual search tasks that underlie many HCI applications in domains such as Extended Reality or auditory interfaces.

\subsection{Resource-Rational Formulation}

We frame audiovisual search as a \textit{resource-rational} decision problem.
Resource rationality, a principle from cognitive science, describes how people approximate rational behavior under cognitive and physical constraints~\cite{Lieder_Griffiths_2020}.
In theory, one would imagine an ideal observer who performs an exhaustive optimal search, \ie considers every possible location, updates probabilities exactly, and searches until certainty is achieved.
In practice, people do not behave this way: they stop early, take shortcuts, and trade off accuracy against the time, effort, and risks involved. 
Thinking in terms of resource rationality allows us to capture these bounded strategies that trade off accuracy or utility~($U$) with cost~($C$), which is especially important in embodied audiovisual search, where each head turn, step forward, or hesitation comes with a real cost of time and physical effort.

The agent's decision problem is therefore to maximize net utility, expressed as:
\begin{equation}
\pi^* = \arg\max_\pi \; \mathbb{E}\big[\sum_{t=0}^{T}\gamma^t( U(b_t, a_t) - C(a_t)) \;\big],
\end{equation}
where $\pi$ is a policy mapping each belief state $b_t$ to actions $a_t$. 
\rv{The index $t$ denotes the decision step, and $T$ is the episode horizon. 
The term $\gamma\in[0,1]$ is a discount factor that controls how future costs and rewards are weighted over time.}
The function $U(b_t,a_t)$ denotes the utility obtained under belief $b_t$ for successfully locating the target, and $C(a_t)$ denotes the cost of performing action $a_t$.
This captures the intuition about human behavior that humans continue searching as long as the extra effort seems worthwhile, but commit to a decision once their uncertainty has been reduced enough that further searching would not justify the extra cost. 
In other words, committing is not about eliminating uncertainty completely, but about deciding that ``good enough'' confidence has been reached given the cost already spent.

Utility~$U$ reflects the benefit of successfully identifying the target source.
At the end of a search task, the agent receives a positive reward if its estimate $(\hat{r},\hat{\theta})$ falls within  tolerance bounds of the true target location, and a penalty otherwise.
The value of committing therefore depends on how concentrated the posterior belief distribution $p(r,\theta|audio,visual,prior)$ has become. 
When the distribution is broad and uncertain, committing is risky; when it is sharp and unimodal, the expected utility of committing increases. 

Embodied actions play an important role in shaping this expected utility.
A head turn changes the geometry of interaural time differences cues, reducing front-back confusion and increasing confidence in azimuth.
Forward movement alters not only range but also the relative angle to the target, producing predictable shifts in both azimuthal and radial evidence.
These changes help distinguish nearby targets from distant distractors and disambiguate sources that were previously overlapped. 
Even small reorientations can transform ambiguous evidence into decisive evidence, effectively sharpening the posterior and raising the expected reward of committing. 
Actions are not merely costs to be minimized but also investments that can increase the eventual payoff of committing.

Cost~$C$ captures not only external costs but also internal constraints of human embodiment.
Each embodied action carries a cost, reflecting the metabolic and temporal expenses of moving, as well as the risk of collisions in cluttered environments. In our formulation:
\begin{itemize}
    \item Head turns incur a small penalty, modeling the energetic cost of reorienting to gather new binaural information.
    \item Locomotion (forward movement) incurs a larger penalty, capturing effort and potential risk of collision with distractors.
    \item Collisions produce a severe penalty, mirroring failure in embodied interaction.
    \item Passive waiting carries a time cost of delaying task completion and preventing timely responses when speed may be critical.
\end{itemize}
In this way, the cost structure encodes both the external environment and the internal limits of human cognition and physiology, ensuring that the model captures the real trade-offs people face when deciding whether additional exploration is worth the expense.
\rv{The specific parameter values used in our study are listed in Appendix~\ref{sec:model-parameters}.}

\subsection{Multisensory Perception and Belief Update} \label{sec:multisensory-belief}
Perception in our model emerges from combining auditory and visual cues.
Auditory cues provide continuous, omnidirectional evidence that orients the agent even when the target is outside the field of view or occluded, while visual cues provide precise, feature-rich confirmation but only when the target is directly visible. 

To represent spatial uncertainty, the agent maintains its beliefs in an egocentric polar grid world.
We discretize space into a polar grid $R \times \Theta$, where $R$ is a set of radial distances and $\Theta$ a set of azimuth angles. 
\rv{This grid has a resolution of 1~m in range and 1\textdegree{} in azimuth, and it defines the space over which the agent computes and updates its posterior.}
At each timestep, the agent maintains a posterior belief distribution $b_t(r,\theta)$ over this grid, expressing how likely it is that the target lies at each \rv{polar coordinate} $(r,\theta)$ \rv{relative to the agent}. 
\rv{As shown in \autoref{fig:model-architecture},} the posterior is \rv{updated} by combining new sensory likelihoods~(auditory and visual) with the prior belief from the previous timestep, \rv{allowing accumulated evidence and uncertainty to propagate over time}.

\begin{figure*}[h]
    \centering
    \includegraphics[width=0.9\linewidth]{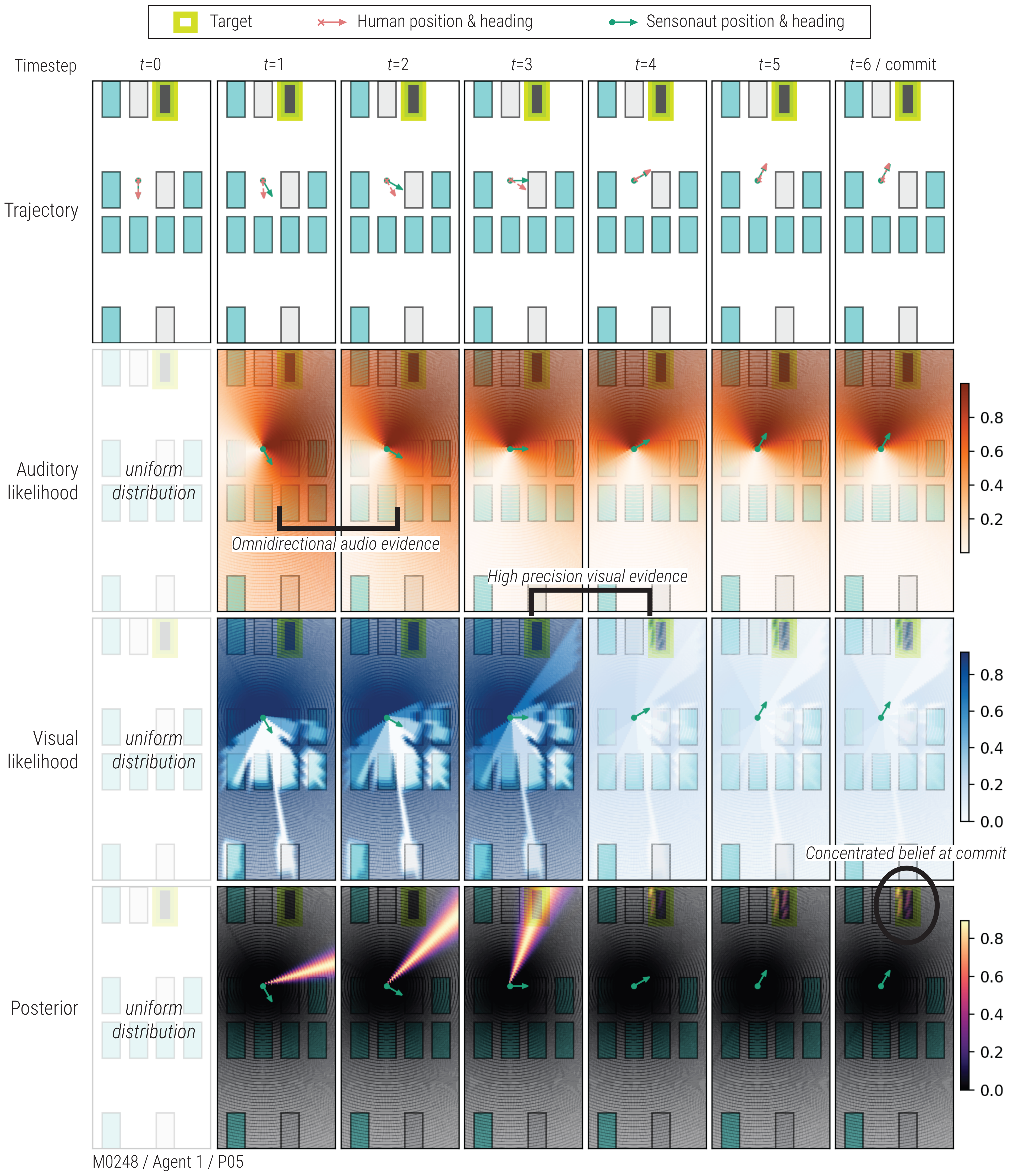}
    \vspace{-0.5em}
    \caption{Example trajectory and belief update of \system{} across timesteps. The top row shows the agent’s position and heading (green solid arrow) and a human participant's position and heading~(pink dashed arrow) in the map's grid view, with a hidden target (green-highlighted box). 
    The second and third rows show the evolving auditory and visual likelihoods: audition offers omnidirectional evidence while vision provides high precision but view-dependent and occlusion-limited evidence. 
    The bottom row shows the posterior belief (which serves as the prior for the next timestep), integrating both modalities and prior. Over time, embodied actions such as head turns reshape the likelihoods, reduce ambiguity, and concentrate belief mass on the true target location (circled).
    }
    \vspace{-1em}
    \Description{Example trajectory and belief update of \system{} across timesteps. The top row shows the agent’s position and heading (green solid arrow) and a human participant's position and heading~(pink dashed arrow) in the map's grid view, with a hidden target (green-highlighted box). 
    The second and third rows show the evolving auditory and visual likelihoods: audition offers omnidirectional evidence while vision provides high precision but view-dependent and occlusion-limited evidence. 
    The bottom row shows the posterior belief (which serves as the prior for the next timestep), integrating both modalities and prior. Over time, embodied actions such as head turns reshape the likelihoods, reduce ambiguity, and concentrate belief mass on the true target location (circled).}
    \label{fig:belief-update}
\end{figure*}

\autoref{fig:belief-update} illustrates an example belief update. 
In the top row, the \rv{green arrow shows the agent and its heading, while} the pink arrow shows an example human trajectory.
Blue, white, and black rectangles denote objects in the scene, with the bold light green outline marking the target.
The three rows show \rv{audio-based belief, visual-based belief,} and posterior that integrates audiovisual evidence with the prior, which is the previous step's posterior.

\subsubsection{Auditory likelihood}
Auditory cues provide continuous, omnidirectional evidence in audiovisual search: they are always available regardless of where the target lies, even when it is behind the listener or occluded from view. 
We represent this contribution as an auditory likelihood $\rv{\mathcal{L}}_{\text{audio},t}(r,\theta)$ on the same polar grid $R \times \Theta$ at timestep $t$.

We base our model on interaural time difference~(ITD) cues, which dominate human judgments of horizontal~(azimuthal) direction for broadband sounds~\cite{shinn2000tori}.
\rv{We model the auditory likelihood by assuming that the conditional probability distribution of the ITD given $\theta$ follows a Gaussian distribution:}
\begin{equation}
\mathcal{L}_{\text{audio},t}(\theta)
= \frac{1}{\sqrt{2\pi\sigma_{\text{ITD}}^{2}}}
\exp\!\left(
  -\frac{({itd}_t - \widehat{\text{ITD}}(\theta))^{2}}{2\sigma_{\text{ITD}}^{2}}
\right)
\end{equation}
\noindent{where $\sigma_{\text{ITD}}$ is the ITD observation noise, and $\widehat{\text{ITD}}(\theta)$ is the predicted ITD under azimuth hypothesis $\theta$.}
$\widehat{\text{ITD}}(\theta)$ is computed on the azimuthal grid using Woodworth’s spherical head model~\cite{woodworth1954experimental}:
\begin{equation}
\widehat{\text{ITD}}(\theta) = \frac{r_{head}\theta + r_{head}\sin\theta}{c},    
\end{equation}
where $r_{head}$ is the average human head radius ($\approx8.75 \text{cm}$), $\theta$ is the azimuth of the sound relative to the head’s forward direction, and $c$ is the speed of sound ($343 \text{m/s}$). 
This function produces a lookup table of ITD predictions across all azimuth hypotheses.
The observed ITD $itd_t$ is a single value available to the agent at each timestep $t$,
simulated by taking the true ITD at the target azimuth $\widehat{\text{ITD}}(\theta^*)$ and perturbing it with Gaussian noise~$\epsilon_{\text{ITD}}$: $itd_t=\widehat{\text{ITD}}(\theta^*)+\epsilon_{\text{ITD}}$,
$\epsilon_{\text{ITD}}\sim\mathcal{N}(0,\sigma^2_{\text{ITD}})$,
producing a noisy but unbiased sample for robust training.

Because ITD is primarily an azimuthal cue and conveys negligible radial information, we model the auditory likelihood as independent of range:
\begin{equation}
\mathcal{L}_{\text{audio},t}(r,\theta) = \mathcal{L}_{\text{audio},t}(\theta), \quad \forall r.    
\end{equation}
This reflects the fact that ITD provides robust directional information but weak radial (distance) sensitivity~\cite{woodworth1954experimental,shinn2000tori,blauert1997spatial}.
Disambiguating range is instead achieved through embodied action and visual observations.

\subsubsection{Visual likelihood}
Visual cues provide complementary, high precision evidence by anchoring probability mass to objects that can be directly seen and by ruling out regions that have already been checked visually.
We represent this contribution as a visual likelihood $\mathcal{L}_{\text{visual},t}(r,\theta)$ on the egocentric polar grid $R\times\Theta$ at timestep $t$, where each pair $(r_i,\theta_i)$ indexes a discrete range–azimuth bin.

We first construct the visual evidence map $E_{\text{visual},t}(r,\theta)$.
At each timestep, the agent has a heading angle $\psi_t$.
Objects whose bearing $\theta$ falls within a $110^\circ$ field of view centered on $\psi_t$ are considered in view.
To handle occlusions, among objects whose bearings differ by less than $5^\circ$, only the nearest is treated as visible, which yields a set of visible polar grid cells $\{(r_i,\theta_i)\}$.
Each visible object contributes a peak at its grid cell $(r_i,\theta_i)$, weighted by feature similarity.
These contributions are summed and normalized by the number of visible objects so that total mass remains bounded.

Vision also provides negative evidence when regions are inspected and found empty.
Along each field-of-view ray, we march outward from the agent until the first occluder is encountered.
All pre-occluder cells are marked as line-of-sight (LOS) empty and multiplied by a decay factor $(1-\delta)$.
If the first visible surface along a ray does not match the target’s appearance, the same discount is applied to all cells up to that surface, which reduces confusion from distractors.
After these LOS and surface discounts, the map is renormalized to obtain $E_{\text{visual},t}(r,\theta)$.

Because visual information is transient and depends on viewpoint, we accumulate it over time in the agent’s egocentric frame.
Let $\Delta\psi_t$ be the change in heading between timesteps $t-1$ and $t$.
Before combining with the new evidence, we rotate the previous visual likelihood by shifting it along the azimuth dimension with wrap-around in $\theta$: $\mathcal{L}_{\text{visual},t-1}\big(r,\theta - \Delta\psi_t\big)$.
We then blend the shifted previous map with the current evidence using an exponential moving average:
\begin{equation}
\begin{split}
\tilde{\mathcal{L}}_{\text{visual},t}(r,\theta)~=~& (1-\lambda_{\text{visual}})
\,
\mathcal{L}_{\text{visual},t-1}(r,\theta - \Delta\psi_t) \\
& +
\lambda_{\text{visual}}
\,
E_{\text{visual},t}(r,\theta),
\end{split}
\end{equation}
and normalize over $R\times\Theta$ to obtain
\[
\mathcal{L}_{\text{visual},t}(r,\theta)
=
\frac{\tilde{\mathcal{L}}_{\text{visual},t}(r,\theta)}
{\sum_{r,\theta}\tilde{\mathcal{L}}_{\text{visual},t}(r,\theta)}.
\]
This produces a head-centered, temporally smoothed visual likelihood that places mass on visible, feature-consistent objects and de-emphasizes regions that have been seen and found empty.
\autoref{fig:belief-update} illustrates how the visual likelihood sharpens around the target when it first enters the field of view and how excluded regions stay suppressed over time.

\subsubsection{Belief update}
At each timestep, the agent maintains a posterior belief $b_t(r,\theta)$ over candidate target locations on the polar grid.
We assume a uniform prior over locations and that auditory and visual observations are conditionally independent given the true target position.
Under these assumptions, we first calculate the instantaneous joint likelihood $\mathcal{L}_{joint,t}(r,\theta)$ by taking the product of the auditory and visual likelihoods:
\begin{equation}
    \mathcal{L}_{joint,t}(r,\theta) = \mathcal{L}_{\text{audio},t}(r,\theta)\mathcal{L}_{\text{visual},t}(r,\theta).
\end{equation}
Taking logs yields an additive form:
\begin{equation}
    \log{\mathcal{L}_{joint,t}(r,\theta)} = \log\mathcal{L}_{\text{audio},t}(r,\theta) + \log\mathcal{L}_{\text{visual},t}(r,\theta).
\end{equation}
In log-space, stronger or more consistent cues naturally dominate the posterior, while weaker or inconsistent cues contribute less, reproducing the reliability-based cue integration observed in 
human multisensory perception~\cite{Ernst_Banks_2002,Alais_Burr_2004}.

To accumulate evidence across timesteps without overcommitting to early observations, we apply a leaky temporal update in log space:
\begin{equation}
\log b_t(r,\theta)=(1-\alpha)\,\log b_{t-1}(r,\theta) + \alpha \,\mathcal{L}_{joint,t}(r,\theta) ,
\end{equation}
where $\alpha$ controls the influence of new evidence.
This update allows the posterior to remain multimodal, supporting competing hypotheses from distractors while reinforcing consistent cues.
The resulting belief evolves smoothly over time and reflects both the stability of prior evidence and the reliability of current observations.

\subsection{Decision Making Under Uncertain Multisensory Perception}
Once the agent forms a posterior belief $b_t(r,\theta)$, it must decide how to act on it. 
This decision process is shaped by two fundamental challenges: 
(1) the agent never directly observes the true state of the world, but only receives partial and noisy auditory–visual cues; and 
(2) every embodied action carries costs in time, effort, or risk. 
As a result, the agent faces a resource-rational problem: it must weigh the expected utility of committing to a decision against the cost and potential benefit of further exploration.

When the posterior is sharp and concentrated on one hypothesis, the utility of committing is high relative to the cost of further exploration, and the agent may terminate the search. 
In the posterior plots of \autoref{fig:belief-update}, as a result of head turns, the probability over the target location starts to become sharper from $t=3$.
However, beliefs are often diffuse or multimodal.
Cluttered scenes may contain distractors that resemble the target in appearance or location, producing competing peaks in the posterior distribution and lowering the expected utility of committing. 
Instead of discarding these distractors, the model preserves them as alternative hypotheses until additional evidence becomes decisive, mirroring how humans defer commitment under uncertainty, as later shown in Figures~\ref{fig:embodied-actions}c and \ref{fig:errors}. 

In these cases, the agent weighs whether additional embodied actions are worth their cost. 
Head turns are relatively inexpensive and can bring rich information gain from both channels: 
they reshape ITD geometry to resolve front–back ambiguities, and they extend the visual field to reveal objects that were previously outside of view.  
Forward movements are more costly, but sometimes essential as they alter both range and azimuth, separating two ambiguous objects aligned along the same line of sight or auditory bearing. 
They also allow the agent to move past occluders, bringing hidden objects into view for confirmation.  
Staying does not incur physical efforts, but delays progress.

In this way, actions emerge as resource-rational choices: the agent acts not only to exploit its current best estimate, 
but also to strategically transform uncertain or ambiguous beliefs into more decisive ones. 
The decision to commit or continue searching is thus shaped jointly by the posterior belief and the costs of embodied action.

\subsection{POMDP Formulation}
We formalize this control problem as a partially observable Markov decision process (POMDP) 
$\mathcal{M}=\langle \mathcal{S},\mathcal{A},\mathcal{O},T,\Omega,R,\gamma\rangle$,
which specifies how hidden states, actions, and noisy observations interact and how beliefs evolve over time.

\begin{itemize}
    \item \textbf{State $\mathcal{S}$:} target location in egocentric polar coordinates, agent's position and heading in world coordinates, and static scene layout with distractors and occluders. This state is hidden from the agent and cannot be observed directly.
    \item \textbf{Actions $\mathcal{A}$:} {\texttt{turn\_left}, \texttt{turn\_right}, \texttt{move\_forward}, \texttt{stay}, \texttt{commit}}. These represent the embodied actions of reorienting, moving forward, waiting, or deciding.
    These are discrete actions, with turn actions using a fixed turn angle $\theta_{turn}$ and move action using a fixed step size $d_{step}$.
    \item \textbf{Observations $\mathcal{O}$:} 
    At each timestep, the agent receives a cognitive state composed of: (i) the full posterior belief distribution $b_t(r,\theta)$ over egocentric polar coordinates, (ii) summary features including the maximum a posteriori~(MAP) estimate and uncertainty in $r$ and $\theta$, (iii) the last actions taken, and (iv) the elapsed time.  
    The agent does not observe raw auditory or visual cues directly. 
    Instead, these cues are processed at the perceptual level of the model and integrated into the posterior belief. 
    This design mirrors human cognition: decision making operates on an uncertainty-aware perceptual state, rather than directly on sensory measurements.
    \item \textbf{Transition $T$:} At the \textit{world level}, actions deterministically change the agent's embodiement: turns rotate the heading, forward steps shift the position by a fixed distance, stay leaves the state unchanged, and commit transitions to an absorbing terminal state.
    At the \textit{belief level}, these embodied changes transform the posterior. A turn rotates the belief distribution along azimuth, aligning it with the new head-centered reference frame. 
    A forward step shifts probability mass inward along the heading direction, reflecting reduced distance to objects ahead.
    Staying does not alter the agent’s physical state but reinforces the most recent observation relative to the prior, gradually sharpening the posterior when the perceptual input remains consistent. 
    Commit or collision terminates the episode. 
    These transitions capture how actions reshape not only the external world but also the information available to the agent through its embodied perspective.
    \item \textbf{Observation model $\Omega$:} the likelihood of receiving $itd_t$ given a candidate azimuth is defined by the Gaussian model around $\widehat{ITD}(\theta)$. Visual likelihoods add localized evidence for visible objects, modulated by feature similarity and occlusion.
    \item \textbf{Reward $R$ ($=U-C$):} 
    Committing correctly yields positive utility, while incorrect commits incur penalties. 
    All actions incur a per-step time cost, reflecting the opportunity cost of delaying task completion. 
    Turns, forward steps, and collisions carry graded costs that reflect time, effort, and risk, while staying has no additional cost. 
    This reward structure encodes the resource-rational trade-off between accuracy and cost
    \item \textbf{Discount factor $\gamma$:} We set $\gamma=0.99$ so that future rewards are slightly discounted relative to immediate ones. This encourages the agent to value quicker resolutions over unnecessarily long search trajectories, while still allowing multi-step exploration when it is expected to reduce uncertainty.
\end{itemize}

\subsection{Parametrization and Policy Learning}
Finding the optimal way to act under uncertainty in a POMDP is intractable in practice, 
so \system{} learns an approximate policy~$\pi(a \mid b)$ for mapping beliefs $b$ to actions $a$ through reinforcement learning.
Using proximal policy optimization (PPO), the agent improves its policy by trial and error, 
guided by rewards for accurate localization and penalties for costly or inefficient actions.
We use Stable Baselines 3 for PPO implementation and provide a list of parameters used in our training \rv{in \autoref{sec:training-details}}.

Through training, the agent develops behaviors of resource-rational audiovisual search: turning to collapse symmetric peaks, moving forward to disambiguate aligned or occluded objects, and committing once beliefs become sharp and unimodal. 
In this way, reinforcement learning links the formal POMDP framework to embodied strategies that resemble human search.

\begin{figure*}[t]
    \centering
    \includegraphics[width=\linewidth]{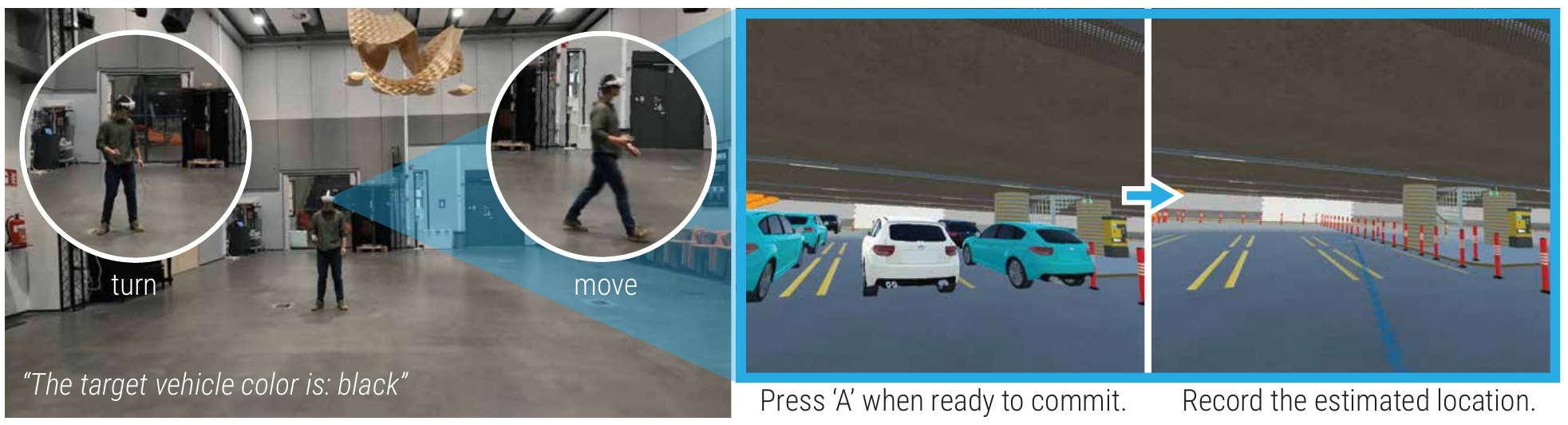}
    \vspace{-0.5em}
    \caption{Data collection setup. Before each trial, participants wearing a Meta Quest 3 saw a text instruction indicating the target car color. 
    During the trial, they physically turned and walked in a VR parking lot to locate the sound-emitting target vehicle. 
    To commit, they pressed the `A' button on the Quest controller, which cleared all cars, and then recorded their estimated target location using a projectile ray selector.
    }
    \vspace{-1em}
    \Description{Experimental setup for collecting human audiovisual search behavior. Participants wore a Meta Quest 3 and physically walked and rotated in a VR parking lot to locate a sound emitting target vehicle. 
    They pressed the `A' button on the Quest controller to commit, which removed all cars, and then recorded their estimated target location using a projectile ray selector.
    Clearing the scene ensured that the estimate reflected the commit moment rather than extra time and observation spent on target selection.}
    \label{fig:data-collection-setup}
\end{figure*}

\section{Data Collection: Human Audiovisual Search}
To compare the model simulation against human behavior, we collected a human audiovisual search dataset in a controlled VR parking lot environment.

\subsection{Methods}
\subsubsection{Participants}
A total of 12 individuals (6 female, 5 male, 1 non-binary; age range 21–33, $M$ = 26, $SD$ = 4.42) were recruited from the local university community.
All participants reported normal or corrected-to-normal vision, hearing, and motor ability. 
Convenience sampling was used, targeting students and staff available for in-person VR studies.  
Participants completed all trials in a single session and were compensated with a €20 gift card.  

\subsubsection{Apparatus}
The study was conducted in a large room where participants could physically walk and rotate~(\autoref{fig:data-collection-setup}). 
The VR setup used a Meta Quest 3 headset and its right controller for input. 
The virtual environment was a Unity-implemented parking lot populated with cars of three colors (blue, black, and white). 
One car was designated as the target, emitting a car beeping sound in loop.
An ambient background noise of the parking garage was played in loop to add realism and auditory noise.
The spectrograms of the car beeping sound and the background noise are provided in the appendix~(\autoref{fig:audio-cue-spectrogram}).  

\paragraph{Auditory setup.}
The spatially oriented format for acoustics (SOFA) \cite{majdak2022spatially} of the head-related transfer function~(HRTF) of a KEMAR~\cite{gardner1994hrft} was used in the Steam Audio engine to spatialize the target and background sounds.
We chose this generic HRTF over individualized HRTF in this experiment as prior work has consistently shown that generic HRTF does not have a significant effect on accuracy in dynamic auditory localization~(\eg~\cite{planinec2023accuracy,riedel2025effect}).

\paragraph{Visual setup.}
The Meta Quest 3 headset has 110 degrees horizontal and 96 degrees vertical field of view with 2,064 $\times$ 2,208 pixels per eye display resolution. 
Participants were instructed to adjust the mechanical lens spacing for their best visual clarity.

\subsubsection{Study Design}
A within-subject experimental design was used, with each participant completing all conditions.  

\paragraph{Independent variables.} The experimental maps varied (a)~\initangle{}: initial angle to target~(front, side, or back), (b)~\numobjs{}: total number of objects in the scene~(5, 7, or 12 cars), and (c)~\numdist{}: number of visual distractors~(0, 2, or 4 cars of the same color as the target), resulting in 27 conditions~($3\times3\times3$).

For controlled map generation, the target object was first assigned to a random spot and color.
The \numobjs{} - 1 remaining cars were then randomly placed in the other slots. 
To control for visual distractors, the colors of the non-target cars were assigned such that exactly \numdist{} - 1 of them matched the target's color.
Lastly, the agent's initial position and heading were sampled randomly until the resulting angle to target fell within the range of the corresponding \initangle{}: front~(-55\textdegree{} to 55\textdegree{}), back~(125\textdegree{} to 235\textdegree{}), or side~(55\textdegree{} to 125\textdegree{} and -125\textdegree{} to -55\textdegree{}).
These ranges are based on the 110\textdegree{} field of view supported by Meta Quest 3.
In this way, the continuous relative angle was treated as a categorical independent variable (\initangle{}) for the experimental conditions and analyses.

\paragraph{Dependent variables.} For each trial, we measured the search time, cumulative head rotation, displacement, and localization success based on the estimated target angle and distance. 
Search time was computed as the duration between the task start and commit event.
Cumulative head rotation was calculated for humans by summing the frame-to-frame changes in yaw, and for our model by deriving the sum from discrete actions (30\textdegree{} per rotation step).
Displacement was calculated as the Euclidean distance between the start and end positions in the Unity XZ plane.
Localization success was recorded as a binary variable indicating whether the final estimate in polar coordinates~($r,\theta$) correctly matched the true target location.
 
\subsubsection{Procedure}
At the beginning of the data collection, participants answered the demographics survey and provided consent of participation.
They were instructed that their task was to find the target vehicle with the target color and a beeping sound as fast and accurately as possible.
Participants were guided through a tutorial (monitored via screen casting) using map variations exclusive to the training phase~(\eg with 2-3 cars and 1 distractor).

Prior to each trial, participants viewed a text instruction indicating the target color.
They used Passthrough mode to safely reorient in the physical room before pressing the \texttt{Meta} button to enter the VR parking lot~(\autoref{fig:data-collection-setup}).
Map configurations were pre-generated based on experimental conditions; all participants experienced the same set of maps in a randomized order. 
During the trial, participants physically navigated by walking and turning, relying on auditory and visual cues to find the target. 
When confident, they pressed \texttt{A} to commit~(\autoref{fig:data-collection-setup} middle).
This action logged the search time and cleared the cars before participants recorded their estimate using a projectile ray selector~(\autoref{fig:data-collection-setup} right).
This mechanism ensured the location estimate relied solely on information gathered prior to the commit decision. 
Participants completed a total of 270 trials (10 trials per condition). 
They were informed that they could take a break at any time and were asked to flag any trials where physical room constraints restricted their movement.

\subsubsection{Data Logging and Processing}
Behavioral logs (position, orientation, and input events) and outcome measures were automatically saved in Unity.
Cumulative yaw rotation was computed as the sum of absolute angular displacements in yaw across each trial.  
Trials where participants pressed a wrong button or where participants reported not being to move further due to the walls were discarded (1.57\% of all trials).  
Outliers with completion times longer than 3 SD from participant mean were removed (1.3\% of trials).

\section{Results: Human vs. \system{}}
We evaluated \system{} on the same set of 270 maps used in the human study.
To account for stochasticity, we ran the model 12 times with randomized map orders, creating 12 agent individuals for comparison with 12 human individuals.
These evaluation maps were not included in the training episodes, ensuring that performance reflects generalization rather than memorization. 
In the following, we compare our model simulation to human behavior across accuracy, search time, head turns, and displacement, highlighting both the similarities that support our modeling assumptions and the mismatches that provide further insights into embodied audiovisual search.

\subsection{Overview of Human Search Behavior}
We analyzed the effects of \initangle{}~(front, side, back), \numobjs{}~(5, 7, 12), and \numdist{}~(0, 2, 4) on dependent variables: accuracy, search time, head turns, and displacement.
Unless noted otherwise, results are from ANOVAs with Tukey-corrected post-hoc comparisons.
To account for minor data imbalances caused by outlier removal ($\sim$3\%), we report Estimated Marginal Means~(EMM).
\subsubsection{Accuracy}
Logistic regression showed no significant main effects or interactions of \initangle{}, \numobjs{}, or \numdist{} on accuracy (all $p>.10$). 
Accuracy remained uniformly high~(94.7\%) across conditions.

\subsubsection{Search time}
There were significant main effects of \initangle{} ($F_{2,3162}=30.84,p<.001,\eta^{2}=.020$), \numobjs{} ($F_{2,3162}=83.24,\\p<.001,\eta^{2}=.004$) and \numdist{}~($F_{2,3162}=51.72,p<.001,\\\eta^2={.050}$), as well as a two-way interaction: \numdist{}$\times$\initangle{} ($F_{4,3162}=2.99, p<.05,\eta^2=.003$). 
The 3-way interaction was also significant ($F_{8,3162}=4.42,p<.001,\eta^2=.010$).
Finding targets at the back~($M = 4.30$~s, $SE = 0.12$~s) was significantly slower than the side~($M = 3.42$~s, $SE = 0.12$~s): $\Delta M = 0.88$~s, $p<.001$; and front~($M = 3.04$~s, $SE = 0.12$~s): $\Delta M = 1.26$~s, $p<.001$.
For \numdist{}, search time increased monotonically with more distractors: 2 vs. 0 distractors~($\Delta M = 0.90$~s, $p<.001$), 4 vs. 0 distractors~($\Delta M = 2.11$~s, $p<.001$), and 4 vs. 2 distractors~($\Delta M = 1.22$~s, $p<.001$).


\subsubsection{Head turns}
The ANOVA revealed significant main effects of \numdist{}, $F_{2, 3162}=36.79, p<.001, \eta^{2}=.023$, and \initangle{}, $F_{2, 3162}=241.94,p<.001, \eta^{2}=.153$. There were also significant interactions between \numobjs{} and \numdist{}, $F_{4, 3162}=3.86, p=.004,\eta^{2}=.050$, and between \numdist{} and \initangle{}, $F_{4, 3162}=3.24, p=.011, \eta^{2}=.004$.
Head turns increased with \numdist{}: 2 vs. 0 distractors ($\Delta M=21.49$, $p<.001$), 4 vs. 0 distractors ($\Delta M=42.16$, $p<.001$), and 4 vs. 2 distractors ($\Delta M=20.67$, $p<.001$).
\initangle{} had a large effect: side targets required more head rotation than front targets ($\Delta M=35.84$, $p<.001$), and back targets required substantially more head rotation than both front ($\Delta M=106.22$, $p<.001$) and side targets ($\Delta M=70.38$, $p<.001$), indicating that participants turned their heads the most when searching behind them. 

\subsubsection{Displacement}
The ANOVA revealed significant main effects of \numobjs{}, $F_{2, 3162}=5.53, p=.004, \eta^{2}=.003$, \numdist{}, $F_{2, 3162}=69.49, p<.001, \eta^{2}=.041$, and angle, $F_{2, 3162}=31.78, p<.001, \eta^{2}=.019$. There was also a significant interaction between \numobjs{} and \numdist{}, $F_{4, 3162}=2.64, p=.032, \eta^{2}=.003$, and a significant three way interaction among \numobjs{}, \numdist{}, and \initangle{}, $F_{8, 3162} = 4.55, p<.001, \eta^{2}=.011$.
Displacement increases monotonically with the number of distractors: 2 vs 0 ($\Delta M = 0.33$, $p<.001$), 4 vs 0 ($\Delta M = 0.62$, $p<.001$), and 4 vs 2 ($\Delta M = 0.29$, $p < .001$). Participants traveled farther from the start before committing to a target.
Side targets require more walking than front targets ($\Delta M = 0.16$, $p = .006$), and back targets require more walking than both front ($\Delta M = 0.42$, $p < .001$) and side targets ($\Delta M=0.26$, $p < .001$). 

\begin{figure}[t]
    \centering
    \includegraphics[width=\linewidth]{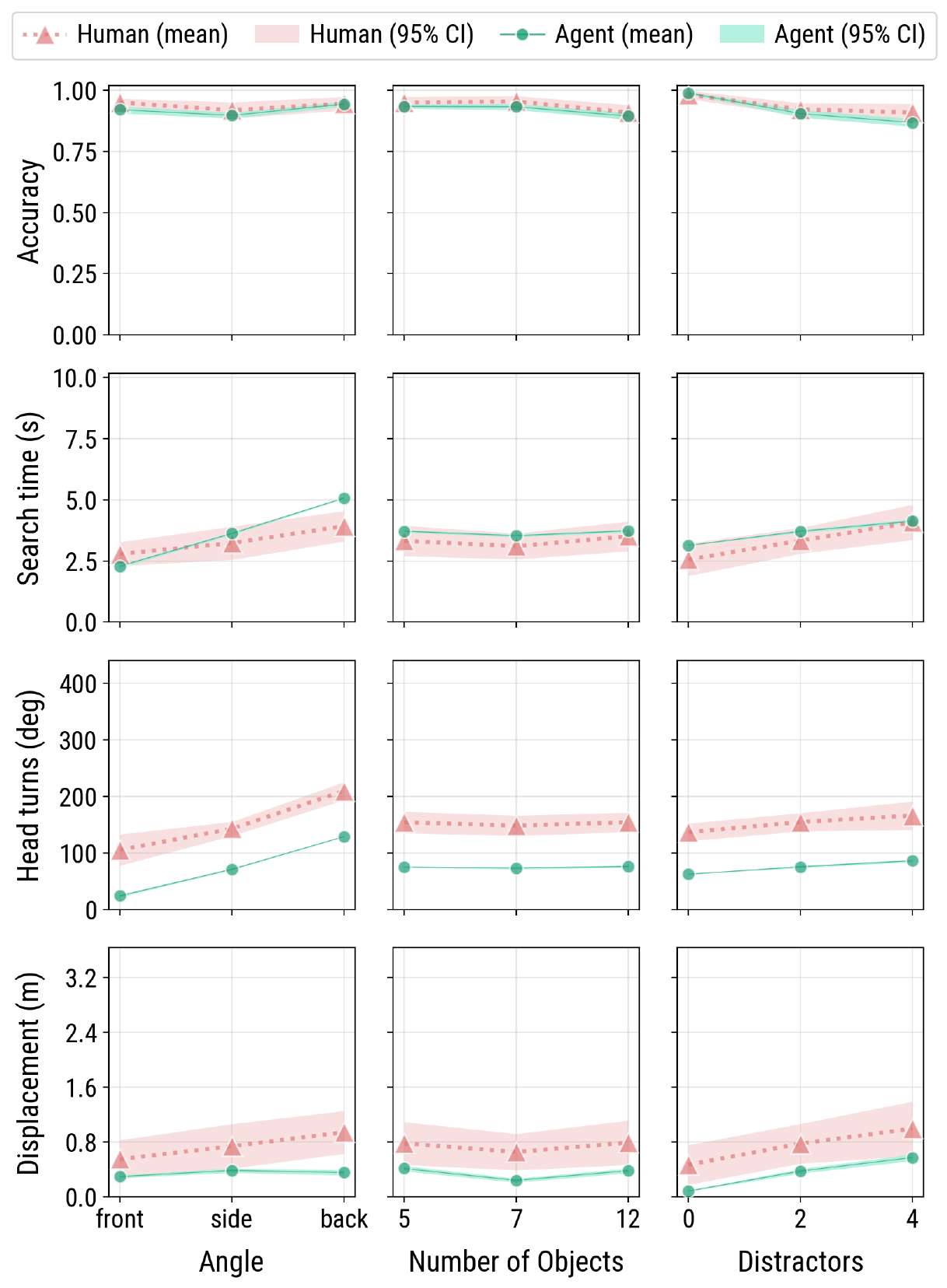}
    \vspace{-0.5em}
    \caption{Human and \system{} performance across initial target angle, number of objects, and number of distractors. 
    Each row shows mean accuracy, search time, head turns, and displacement, with shaded regions indicating 95\% confidence intervals. 
    Both human and agent show longer search times, more head turns, and more locomotion when the target starts to the side or behind, or with more distractors.
    }
    \vspace{-1em}
    \Description{Effects of angle, number of objects, and distractors on search time, accumulated head turns, and distance.}
    \label{fig:tendency}
\end{figure}

\subsubsection{Summary}
Across measures, humans show a selective response to task difficulty:
\begin{itemize}
    \item \initangle{} is the dominant driver (the hardest in the back) for time, displacement, and especially head turns.
    \item \numdist{} increase time and error, but their effect is strongest when the view angle is unfavorable; there is no general slowdown.
    \item \numobjs{} has small main effects but matters via interactions.
    \item Accuracy remains high, suggesting effort is invested primarily to maintain performance rather than to chase marginal accuracy gains.
\end{itemize}

\begin{figure}[t]
    \centering
    \includegraphics[width=\linewidth]{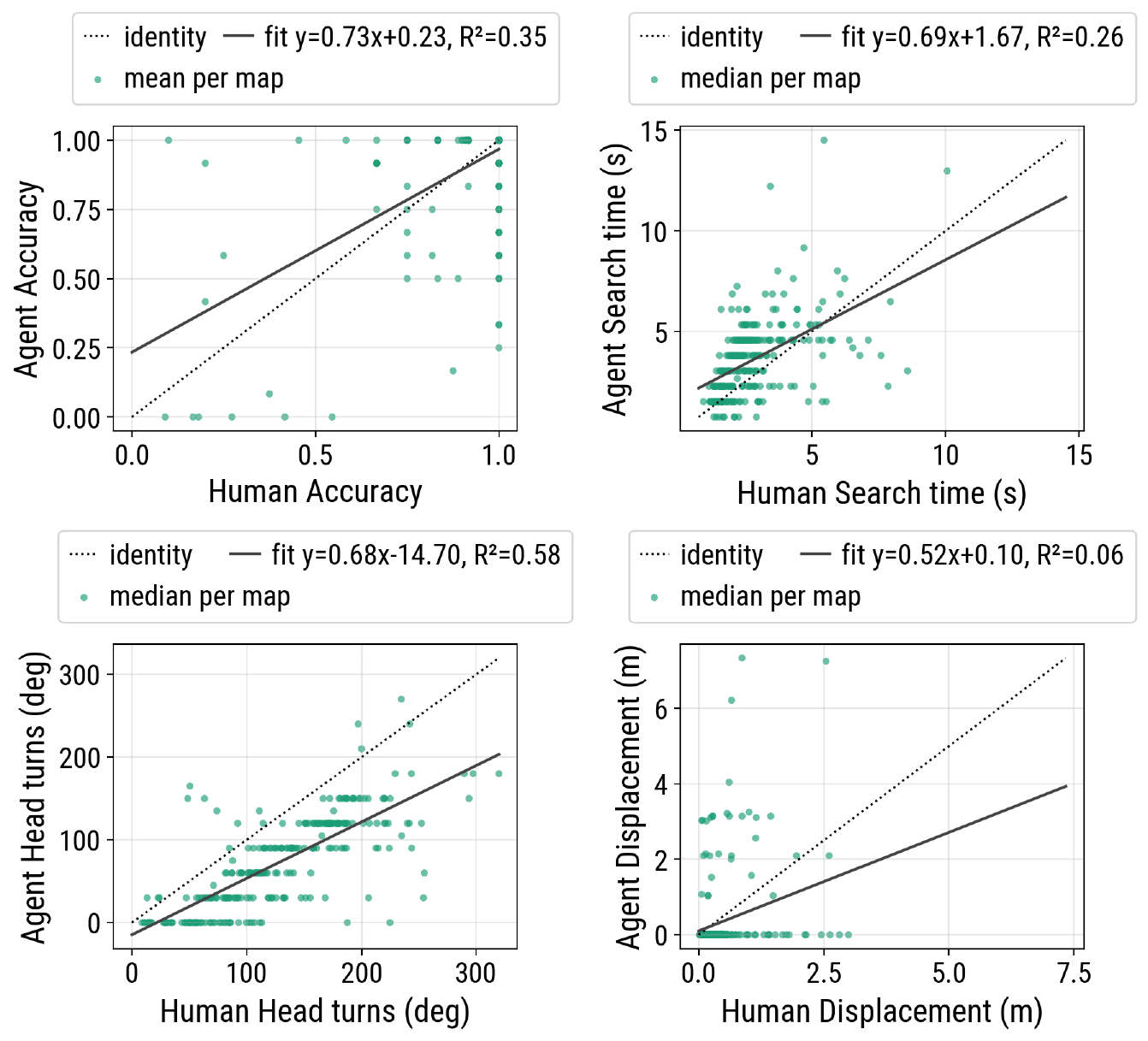}
    \vspace{-0.5em}
    \caption{
    Correlation between human and \system{} on accuracy, search time, cumulative head rotation, and displacement. 
    \system{} reproduces map-level patterns in accuracy, search time, and head turns, with weaker correspondence in displacement, further analyzed in Section~\ref{sec:locomotion-misalignment}.}
    \Description{Overview of model accuracy across three variables: (a) search time, (b) head rotation, (c) locomotion distance. \system{} shows a good match with tendencies in search time and head rotation, but undershoots locomotion distance.}
    \label{fig:correlation}
\end{figure}

\subsection{Performance}
\autoref{fig:tendency} summarizes human and \system{} performance on the four measures of accuracy, search time, accumulated head turns, and displacement across \initangle{}, \numobjs{}, and \numdist{}. 
Across all measures, both humans and the model reveal the same systematic patterns, showing that the model generalizes the central effects of audiovisual search.

Accuracy remains high overall but drops slightly for the side angles and in the presence of distractors, with model closely tracking human performance. 
Search times increase when targets are behind or when distractors are present, reflecting the added difficulty of disambiguating cues under these conditions.
Head turns also rise under these conditions, highlighting how both humans and the model rely on reorientation to resolve ambiguity. 
Humans, however, show larger and more variable head turns, consistent with their ability to make turns of varying magnitude compared to the model’s fixed 30\textdegree{} steps. 
Displacement remains relatively low, with the model showing an even stronger preference than humans to minimize walking, reflecting its explicit action cost.

These patterns demonstrate that our model produces the key embodied strategies of audiovisual search, such as relying primarily on head turns, increasing effort when uncertainty is high, and maintaining high accuracy while minimizing costly movements.

\begin{figure*}[t]
    \centering
    \includegraphics[width=\linewidth]{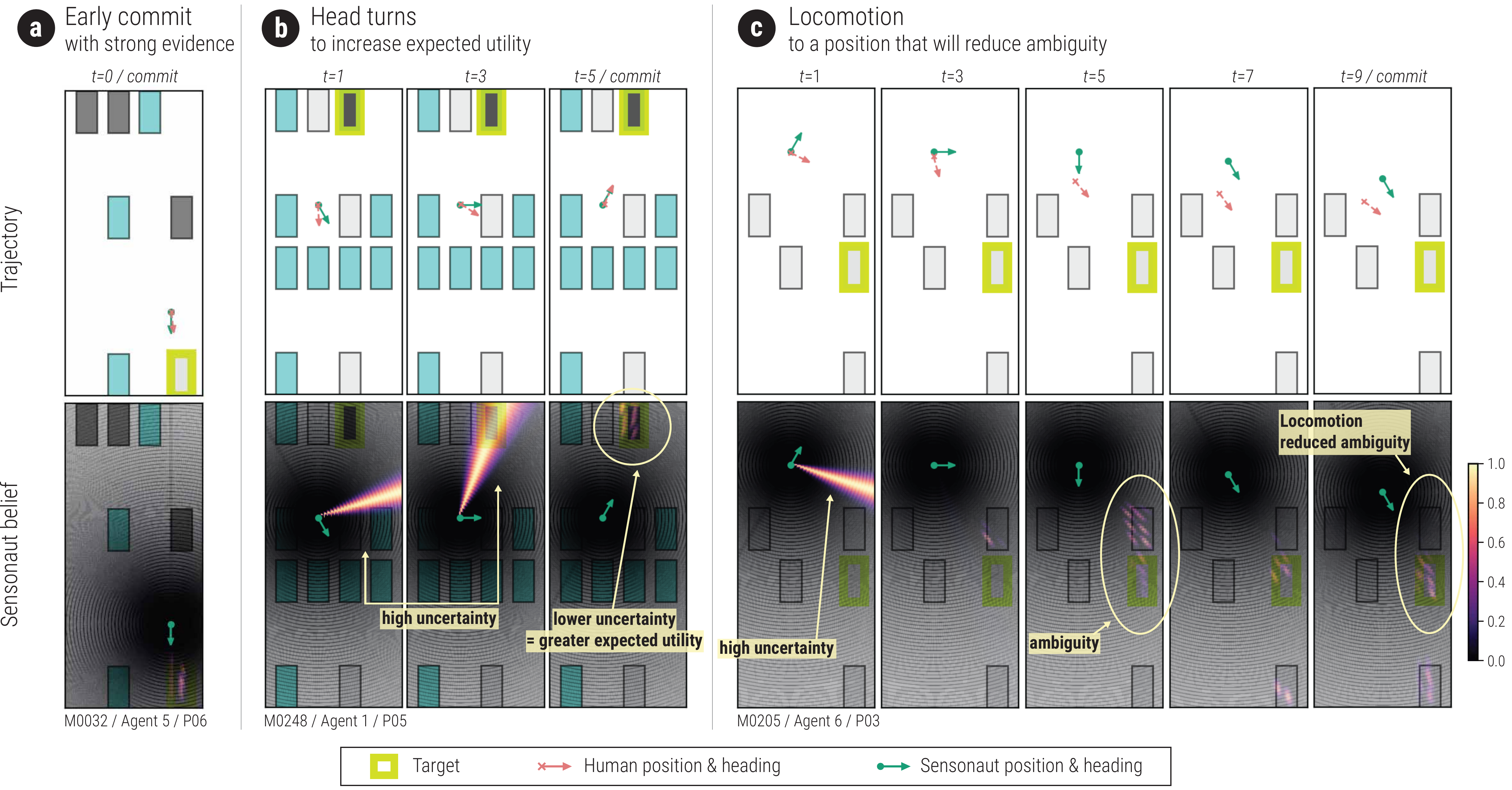}
    \vspace{-0.5em}
    \caption{
    Examples of human-like search strategies produced by \system{}. 
    \textbf{(a)~Early commit with strong evidence.} Both human and agent commit immediately when the initial audiovisual evidence heavily favors one candidate. 
    \textbf{(b)~Head turns to increase expected utility.} When the initial posterior is ambiguous, \system{} performs incremental head turns that sharply reduce uncertainty, matching human behavior.
    \textbf{(c)~Locomotion to a position that will reduce ambiguity.} When rotation alone cannot resolve uncertainty, both humans and \system{} move to positions that provide a clear discriminative cue. In this example, the agent turns toward the unexplored region behind it and then moves toward the midpoint between the confusing objects to obtain a left-right contrast that collapses the posterior.
    }
    \vspace{-1em}
    \Description{Search strategies exemplified through trajectories of human participants and \system{}. The square indicates the initial starting position with the arrow pointing at the human or agent's initial heading. The trajectory is drawn in line. The end point of the human or agent is marked with a cross symbol $\times$, and their committed estimate is marked with $\star$. The target vehicle is indicated with a thick lightgreen border.}
    \label{fig:embodied-actions}
\end{figure*}

\subsection{Correlation}
To systematically compare human and \system{} behavior, we computed the
correlation between map-level medians of human and model predictions~(\autoref{fig:correlation}). 
Comparing medians captures the central tendency of human behavior on each map while smoothing over participant-specific variability.
This provides a stable basis for evaluating whether the model reproduces systematic patterns of embodied audiovisual search across environments.

The fits reveal that the model explains 35\% of the variance in accuracy~($R^2=0.35$), 26\% in search time~($R^2=0.26$), 58\% in head turns~($R^2=0.58$), and 6\% in displacement~($R^2=0.06$).
Accuracy aligns well, showing that the model can predict human errors. 
Search times also align, with the model capturing the trend of longer searches in 
harder conditions, though it tends to be slightly slower than humans. This reflects the 
model’s conservative policy of prolonging exploration until uncertainty is sufficiently 
reduced.  

The strongest match is in head turns, where \system{} captures how participants relied on rotation as a primary, low-cost way of gathering evidence.
Humans could rotate by arbitrary angles and accumulate variable degrees of rotation, whereas \system{} turned only in 30$^\circ$ increments. 
Despite this constraint, the correlation is high, indicating that the model captures the central role of rotation.  
Displacement showed the weakest alignment. 
Participants varied in how far they traveled from the start, whereas the model was more consistent. This high human variability explains the lower predictability for this metric.

While these $R^2$ values are modest, they are consistent with leave-one-out human-to-human correlations: $R^2=0.22$ for search time, $R^2=0.26$ for head turns, and $R^2=0.11$ for displacement. This underscores that perfect fits are unrealistic given the variability of human strategies.
The scatter plots also reveal a stepped pattern in the model’s outputs, a natural consequence of its discretized action space (30\textdegree{} turns, 1~m steps). 
Even so, the model reproduces the main embodied strategies of audiovisual search and provides a principled account of where and why discrepancies arise.

\subsection{Search Strategies}

Our model exhibited embodied search strategies similar to humans, driven by resource-rational decision making.
\autoref{fig:embodied-actions} illustrates three characteristic patterns.

\paragraph{Early commit with strong evidence.}
Both humans and \system{} committed immediately when the audiovisual evidence strongly favored a single candidate.
In these cases, the posterior after the first step of a trial was already sharply peaked (\autoref{fig:embodied-actions}a), resulting in a direct commit without head movement or locomotion.
This behavior maximizes expected utility by prioritizing the commit action over costly exploration when confidence is high.

\paragraph{Head turns to increase expected utility.}
When the initial posterior was misleading or ambiguous---often for targets located behind~($t$=1 in \autoref{fig:embodied-actions}b)---both humans and \system{} used head turns to gather information.
Rotation served a dual purpose.
First, accumulating auditory samples at different angles and visually ruling out incorrect locations corrected the directional estimate by resolving front-back confusion~($t$=3).
Subsequently, bringing the target into the visual field caused the posterior to collapse into a sharp peak~($t$=5). 
This illustrates how the learned policy uses low-cost sensory actions to increase expected utility before committing.


\paragraph{Locomotion to reduce ambiguity.}
In trials where ambiguity persisted even after head turns, both humans and \system{} engaged in locomotion.
At $t$=5 in \autoref{fig:embodied-actions}c, the posterior remained bimodal, split between the true target and a competing distractor.
To resolve this, both agents moved to a position that generated motion parallax, leveraged occlusion geometry, and produced more discriminative audio cues.
The resulting posterior updates successfully ruled out the distractor, causing the posterior to converge on a single candidate, as shown at $t$=7 and $t$=9 in \autoref{fig:embodied-actions}c. 
These locomotion-driven disambiguation emerged naturally from the belief-based policy, allowing the agent to commit with high confidence.



\begin{figure}[h]
    \centering
    \includegraphics[width=0.835\linewidth]{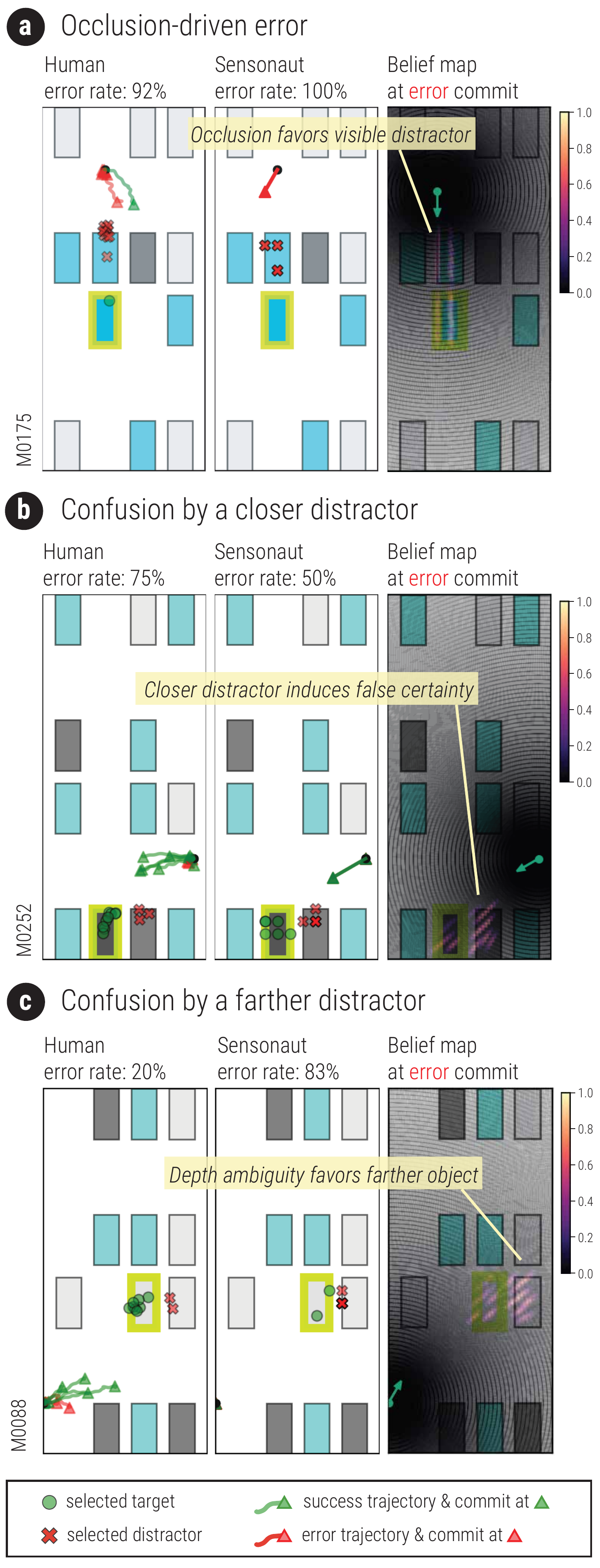}
    \vspace{-0.5em}
    \caption{
    \system{} reproduces human errors arising from sensory uncertainty. Both humans and the model are misled by distractors with similar visual angles but different depths. Belief maps (right) show the cause: perceptual ambiguity leads to false certainty in occlusion (a), proximity (b), and distance (c) scenarios.
    %
    }
    \vspace{-2em}
    \Description{Example of \system{} predicting potential unsuccessful trials. The explainable model of \system{} enables prediction of successful and unsuccessful cases of human audiovisual search, for example, getting confused and distracted by a similar target. }
    \label{fig:errors}
\end{figure}

\subsection{Errors}
To assess whether our model captures not only successful search behavior of humans but also errors, we analyzed all incorrect human trials~(6\% of total trials; manually coded by one author) and grouped them into five categories: occlusion-driven errors, confusion by a closer distractor in a similar direction, confusion by a farther distractor, front-back confusions, and selection mistakes.
\autoref{fig:errors} illustrates three primary error types in which our model reproduces human errors, providing computational explanations for how these errors arise from perceptual ambiguities and scene structures.

\paragraph{Occlusion-driven errors~(35\%)}
The most frequent human errors were due to occlusion, where the target was fully or partially hidden by an intervening object from the starting viewpoint. 
Humans frequently committed to the visible distractor whose appearance and direction matched the auditory cue, such as in both examples shown in \autoref{fig:errors}a. 
Our model exhibited the same error mode, replicating 100\% of occlusion-driven error instances by our human participants. 
The belief plots show that \system{} maintained high uncertainty over the occluded region and selected the distractor because the alternative hypothesis remained weakly supported by the audiovisual evidence available from the starting position.

\paragraph{Confusion by a closer distractor~(28\%)}
Both humans and \system{} committed early to a closer distractor appearing in a similar direction as the target. 
In \autoref{fig:errors}b, the distractor already produced a relatively low-uncertainty posterior mode, even though the true target had not been fully ruled out. 
This made the distractor a locally compelling hypothesis.
Our model reproduced 42.9\% of the human error instances of this type.
In the same map, when either human or the model walked closer to the target, they successfully selected the correct one by reducing ambiguity.

\paragraph{Confusion by a farther distractor (5\%)}
Farther-distraction confusions appeared in layouts where a distant object aligned with the target's direction, creating elongated posterior peaks.
Humans occasionally committed to the farther distractor, and \system{} reproduced this outcome in 30\% of human errors on these maps.
\autoref{fig:errors}c shows a posterior unresolved at commit time, leading to the selection of the distractor despite partially informative cues. 

The remaining error types were less aligned with the scope of our model.
Front-back confusions occurred in human trials when participants committed early to an object in front, even though the true target was behind them. \system{} did not reproduce these errors as the agent consistently performed head-turning actions that resolved the ambiguity before committing.
Selection mistakes reflected failures to recall the target's color or location and are not attributable to audiovisual inference, so we excluded them from evaluation of the model's perceptual behavior.

\begin{figure*}[t]
    \centering
    \includegraphics[width=\linewidth]{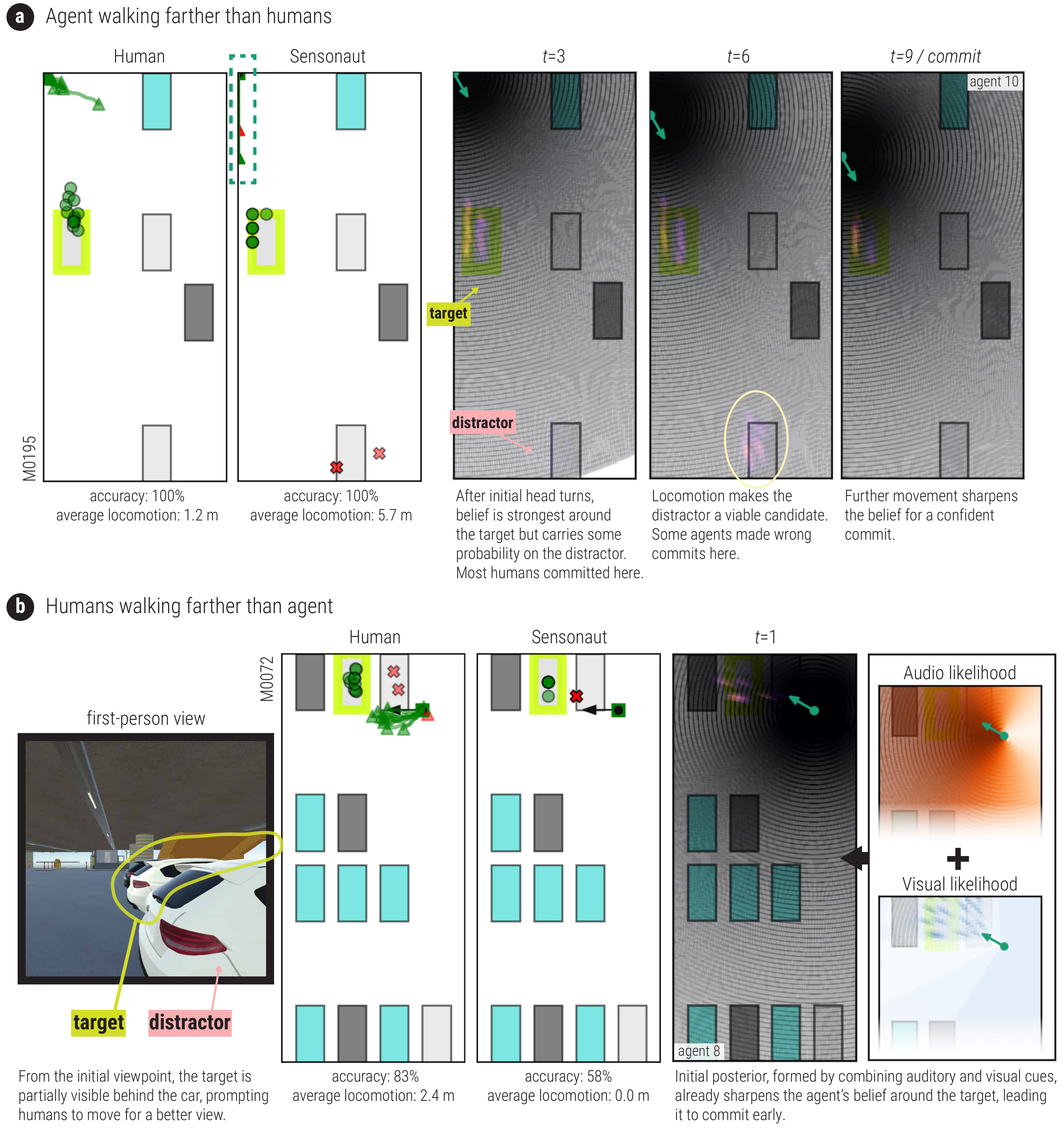}
    \caption{\rv{Cases with the largest human-agent locomotion discrepancies.
    (a)~\system{} walked farther than humans. In such cases, humans tended to commit early once the target became visually plausible, whereas \system{} continued moving to resolve remaining ambiguity, sometimes leading to incorrect commits.
    (b)~Humans walked farther than \system{}. Partial visual information led humans to move for a clearer view, but \system{} stopped early because the initial audiovisual likelihoods already created a sharply peaked posterior that triggered an immediate commit.}}
    \Description{Failure cases of \system{} where it moved less than or farther than human (top). It also could not simulate the behavior when it was completely blocked by occluders (bottom).}
    \label{fig:failure-cases}
\end{figure*}

\subsection{Movement Trajectories} \label{sec:locomotion-misalignment}

We analyzed locomotion discrepancies by identifying maps where the absolute difference between human and model displacement exceeded 1.5~m.
These cases were separated into those where the model overshot locomotion and those where the human walked farther. 
\autoref{fig:failure-cases} depicts the largest deviation observed for each scenario.

\paragraph{\system{} walked farther than humans.}
In several maps, humans committed early despite residual uncertainty in the scene.
In \autoref{fig:failure-cases}a, most participants selected the nearer candidate without taking additional steps, even though the farther object remained a plausible option.
Our model, however, maintained some probability on the farther distractor, prompting it to walk forward to gather more evidence, sometimes leading to incorrect commits.

\paragraph{Humans walked farther than \system{}.}
In other maps, humans took exploratory steps to resolve partial occlusion.
In \autoref{fig:failure-cases}b, they moved to gain a clearer view of a target that was only partially visible.
Our model did not move in these conditions because the initial audiovisual likelihoods already produced a sharply peaked posterior.
This partially arose from the binary treatment of visibility in the current visual likelihood: once any portion of the target was visible, the model treated it as fully visible. 
Moreover, our model could only move in 1~m increments, whereas humans could adjust their position with finer movements.
These led to premature certainty and suppressed locomotion, whereas humans continued to gather evidence when only a fraction of the object was visible.


\section{Discussion}

\rv{Our work aims to advance our} understanding of audiovisual search and provided computational support for it. 
Although audiovisual search is a common task in VR, MR and physical environments, actionable insights into how people solve this task have been lacking. 
In this paper, we introduced a new computational theory that, for the first time, explains and predicts users’ embodied actions. 
Extending previous approaches, such as the Bayesian ideal observer model, our theory suggests that people use head rotation and locomotion to form beliefs that help them distinguish the target. 
Importantly, users account for the costs of such actions and weigh them against the probability of solving the task efficiently. 
Under time pressure, they are compelled to “gamble” and commit to a decision under uncertainty. 

We demonstrated how to implement the theory in a computational model, which we tested against data from a realistic search task. 
\system{} implements resource rationality in a decision-theoretic model (POMDP) that allows us to encode both utilities and costs in the reward function. 
POMDP is a natural fit for the task because it assumes that rewards are delayed (sequentiality) and that actions must be chosen under uncertainty (partial observability). 
However, what POMDP previously lacked was a model of belief formation, which we proposed as probabilistic integration of the two signals. 
When the resulting POMDP is solved using reinforcement learning, the model produces moment-by-moment predictions of search behavior. 

To sum up, its predictions fit three of the four key variables in human data: search time, head rotation, and success rate. 
The model also captures the effects observed in human data as the number of cars and distractors increases, as well as the effect of target direction. 
For movement displacement, however, we observed a discrepancy: humans often take a step at the beginning of a trial, whereas the model avoids this due to the assumed cost. 
In trials involving longer locomotion paths, however, the model provided a better match.
Encouragingly, all of these findings emerged from the same underlying principle. 
Thus, our results provide strong evidence for the viability of resource rationality as a key cognitive mechanism.


\subsection{\rv{Applications and Broader Implications}}
\rv{\system{} can serve as an analytic tool for evaluating in silico how spatial layouts, user interfaces, or assistive systems shape users' ability to localize information.}
Practitioners can train the model in environments that \rv{match their application domain, then use the learned policy to assess which layouts may confuse users, which distractors are most misleading, or which locations produce uncertainty that requires additional time or effort.}

Beyond static simulation, \rv{the model can also serve as an} an objective function for \rv{automated design optimization} or \rv{support} AI assistants \rv{or robots} that \rv{guide users toward more informative viewpoints.}
Furthermore, the model can be executed in a variety of platforms such as Unity3D, which enables applications in MR environments and games. 
To facilitate these applications, we release the model and training code as open source upon publication of this paper. 

\rv{Beyond audiovisual search, the underlying computational structure generalizes to a broader set of HCI problems that involve perception, uncertainty, and active exploration.
Many interactive tasks require users to integrate imperfect cues, evaluate partial visibility, or decide between gathering more information and acting with incomplete knowledge. 
Examples include typing~\cite{shi2024crtypist}, chart reading~\cite{shi2025chartist}, attention switching between Augmented Reality and physical environments~\cite{bai2024heads}, illusory VR interaction~\cite{gonzalez2023sensorimotor,kari2025reality}, and human-robot interaction~\cite{strabala2013toward}.
A belief-based model provides a unified representation of uncertainty across multiple sensory channels and allows simulation of how users gather evidence, where they may become confused, and when premature commitment is likely.
These capabilities suggest braoder use of resource rational models as tools for understanding perception and decision making in interactive systems.
}

\subsection{\rv{Limitations and Future Work}}
\rv{Our analysis revealed potential structural sources of locomotion mismatch.
\system{} currently lacks a representation of human proximity bias, \ie that humans tend to favor the nearer candidate as their first estimate even when a farther candidate remains plausible. This contributed to cases where the agent walk farther than humans.
The visual likelihood also treats visibility in a binary manner for agents, causing premature certainty when only a portion of the target is visible.
In contrast, humans frequently take one or two small exploratory steps to resolve partial occlusion.
These issues were compounded by the agent's coarse action space, which advances one meter per step and separates turning and moving, whereas humans rotate and walk continuously at the same time.
Incorporating finer grained embodied dynamics and graded visibility is a promising next step to improve the effectiveness of our approach.
}


\rv{
\system{} models a single resource-rational policy rather than a distribution of human strategies. 
Thus, our evaluation focuses on \textit{alignment} between human and agent behaviors, similar to prior works on user simulation in HCI~\cite{shi2024crtypist,shi2025chartist,chen2021adaptive}, rather than  classical statistical tests used for null hypothesis testing on the differences between human and agent ``groups'' because \system{} is closer to an individual.
Extending the model with parameterized variations that reflect different human strategies is an important direction.
Although our within subject dataset~(N=12) revealed systematic behavioral patterns, it does not capture the full diversity of human behavior. 
Larger-scale data collection will allow broader validation and support learning of individual differences.
We plan to expand our research in this direction in the future.
Our dataset, model, and data collection software to support this direction are available at \url{https://augmented-perception.org/publications/2026-sensonaut.html}.
}



\section{Conclusion}
Audiovisual search in immersive environments can be understood as resource-rational behavior, where people integrate uncertain auditory and visual cues and strategically move their head and body to reduce ambiguity. We introduced \system{}, a computational model that formalizes this process as a POMDP and uses reinforcement learning to approximate human strategies. The model reproduces key patterns of behavior: high accuracy with small drops under distractors and back angles, longer search times and more head turns under ambiguity, and minimal walking unless necessary.
Crucially, it also reproduces characteristic human failure modes, showing how distractors aligned in visual angle induce false certainty and drive incorrect commitments.
The model explains embodied audiovisual search behavior through belief updating and embodied action. 
By capturing both perceptual integration and action costs, \system{} not only predicts outcomes but also illuminates the processes behind them, offering a foundation for simulation-based design of XR and assistive systems that anticipate user behavior and adapt guidance to human constraints.

\begin{acks}
This work was supported by European Research Council Advanced Grant (no. 101141916), the Research Council of Finland project Subjective Functions (grant 357578), Finnish Center for Artificial Intelligence (grants 328400, 345604, 341763), National Research Foundation of Korea (RS-2023-00223062), and Institute of Information and Communications Technology Planning and Evaluation (RS-2020-II201361).
\end{acks}

\bibliographystyle{ACM-Reference-Format}
\bibliography{references}

\appendix

\section{\rv{Training: Model Architecture and PPO Hyperparameters}} \label{sec:training-details}

\paragraph{Custom Feature Extractor}
We implement a multi-branch feature extractor for dict observations that mixes low-dimensional history vectors with an optional high-dimensional posterior map.

\textbf{Inputs \& branches:}
\begin{itemize}
    \item $\texttt{est\_theta} \in \mathbb{R}^H$: MLP $[H \rightarrow 128 \rightarrow 128]$ with LayerNorm on hidden layer $\rightarrow 128$-D.
    \item $\texttt{theta\_uncertainty} \in \mathbb{R}$: MLP $[1 \rightarrow 32 \rightarrow 32] \rightarrow 32$-D.
    \item $\texttt{est\_r} \in \mathbb{R}^H$: MLP $[H \rightarrow 128 \rightarrow 128]$ with LayerNorm $\rightarrow 128$-D.
    \item $\texttt{r\_uncertainty} \in \mathbb{R}$: MLP $[1 \rightarrow 32 \rightarrow 32] \rightarrow 32$-D.
    \item $\texttt{last\_actions} \in \text{MultiDiscrete}(H, \text{card}=6)$: token embedding (size 8) per timestep (H tokens) $\rightarrow$ flatten $H \times 8$ $\rightarrow$ MLP $[H \cdot 8 \rightarrow 64 \rightarrow 64] \rightarrow 64$-D. Robust to index, one-hot, or flattened one-hot encodings.
    \item $\texttt{posterior} \in \mathbb{R}^{R \cdot \Theta}$ (treated as 2D map): reshape to $(1, R=30, \Theta)$ $\rightarrow$ Conv2d$(1 \rightarrow 16, 3 \times 5, s=(1,2), p=(1,2)) + \text{ReLU}$ $\rightarrow$ Conv2d$(16 \rightarrow 32, 3 \times 3, s=2, p=1) + \text{ReLU}$ $\rightarrow$ AdaptiveAvgPool2d$(1 \times 1)$ $\rightarrow$ LayerNorm(32) $\rightarrow$ Linear$(32 \rightarrow 256) + \text{ReLU} \rightarrow 256$-D.
    \item $\texttt{posterior\_entropy} \in \mathbb{R}$: MLP $[1 \rightarrow 16 \rightarrow 16] \rightarrow 16$-D.
\end{itemize}

\textbf{Output dimensionality:}  
$128 + 32 + 128 + 32 + 64 + 256 + 16 = 656$.  

The extractor concatenates all active branch outputs and returns this feature vector, which is passed to SB3 for policy/value heads.

\paragraph{PPO Policy and Optimization}
We train PPO (Stable-Baselines3) using this extractor as \texttt{features\_extractor\_class}. The actor–critic uses separate MLP heads:

\begin{itemize}
    \item \textbf{Policy head ($\pi$):} $[256, 256, 256]$.
    \item \textbf{Value head ($V$):} $[256, 256, 256]$.
\end{itemize}

\textbf{\rv{Hyperparameters:}}
\begin{itemize}
    \item Algorithm: PPO (SB3)
    \item Policy: MultiInputPolicy (with custom extractor)
    \item Device: NVIDIA GeForce RTX 3090. CUDA 12.9
    \item Learning rate: $3 \times 10^{-4}$
    \item Rollout length: $n\_steps = 4096$
    \item Batch size: $4096$ (full-batch update per rollout)
    \item Gamma: 0.99
    \item GAE lambda: 0.95
    \item n\_epochs: 10
    \item Clip range: 0.2
    \item Training steps: 1,000,000
    \item Seed number: 42
\end{itemize}

\section{Model Parameters} \label{sec:model-parameters}
All model parameters are listed in \autoref{tab:parameter-list}.
\begin{table*}[p]
    \centering
    \caption{Summary of \system{}'s model parameters. Each parameter is listed with its default value and a brief justification based on the task design, perceptual modeling assumptions, or standard practice in deep reinforcement learning~(RL).}
    \label{tab:parameter-list}
    \begin{tabular}{p{3.5cm}cp{11cm}}
    \toprule
        Parameter Name & Value & Description \\
        \midrule
        \rowcolor{gray!20} State and Action & & \\
        History length & 4 & Number of past observations stored in state. \rv{Provides short-term memory for handling partial observability and allows the model to learn the consequences of recent embodied actions. A length of four also follows common partially observable RL practice, such as the 4-frame stack in DQN~\cite{mnih2013playing}.}
        \\
        Turn angle & 30$^\circ$ & Angular increment for each turn action. \rv{Matches mean maximum head angular velocities during walking~\cite{pozzo1990head} and keeps the action space tractable for RL.} \\
        Stride & 1~m & Forward movement step size. \rv{Approximates an average human step distance~\cite{imai2001interaction, pozzo1990head} while ensuring consistent and efficient exploration in grid-based navigation tasks.}\\
        Maximum steps in an episode & 30 & Upper limit on episode length \rv{to encourage efficient search and prevent degenerate long trials.} Common \rv{practice} in RL \rv{navigation} tasks~(\eg~\cite{paul2024multi}), tuned \rv{to the difficulty of our 13 $\times$ 29 search maps}. \\
        
        \midrule
        \rowcolor{gray!20} Reward & &\\
        Task reward & 10 & \rv{Positive} reward for correctly localizing the target. Sets \rv{the scale for other rewards and encourages accuracy.}\\
        Timestep penalty & 0.1 & Small \rv{cost per} timestep to \rv{avoid idling and} encourage \rv{efficient search behavior.} \\
        Forward penalty & 0.3 & Penalty for \rv{locomotion tuned} to reflect \rv{physical} effort costs \rv{and biases toward minimal movement}. \\
        Turn penalty & 0.1 & Penalty for head/body \rv{turning actions to reflect the lower but nonzero embodied cost of rotation. Tuned so that head turns remain cheap but not free.} \\
        Collision penalty & 5.0 & \rv{High p}enalty \rv{to prevent the agent from} colliding with obstacles, \rv{ensuring realistic and safe trajectories}. \\

        \midrule
        \rowcolor{gray!20}Perception & & \\
        Field-of-view & 110$^\circ$ & Agent’s visual field \rv{to approximate the human horizontal field-of-view when using the Meta Quest 3 headset}. 
        \\
        New evidence blending factor $\alpha$ & 0.8 & Weight \rv{controlling how strongly new sensory evidence updates the belief relative to old evidence. Tuned to ensure stable belief updates without oversensitivity.}
        \\
        Visual-audio ratio & 0.7 & Weighting \rv{factor that controls the relative influence of} visual and auditory likelihoods \rv{in the posterior. Based on Bayesian cue-integration principles, where each modality is weighted by its reliability~\cite{Alais_Burr_2004,Ernst_Banks_2002}.}. 
        \\
        ITD noise scale & 30 $\mu$s & Noise scale applied to interaural time difference\rv{~(ITD) cues. Matches known human detection thresholds for ITD discrimination at low frequencies~\cite{grantham1978detectability}.} 
        \\
        Visual evidence blending factor $\lambda_\text{visual}$ & 0.7 & \rv{Relative} weight for integrating new vs. past visual evidence. Tuned \rv{empirically to match clear visual dominance in human behavior.}
        \\
        Visible objects weight & 5.0 & Weight \rv{emphasizing contributions of} visible objects in belief update\rv{s, ensuring strong influence of visual confirmation. Tuned to facilitate realistic audiovisual integration}.
        \\
        \\
        Visual exclusion decay $\delta$ & 0.5 & Decay rate for excluding previously seen objects. Tuned empirically. 
        \\
    \bottomrule
    \end{tabular}

    \vspace{0.8cm}
    
    \includegraphics[width=\linewidth]{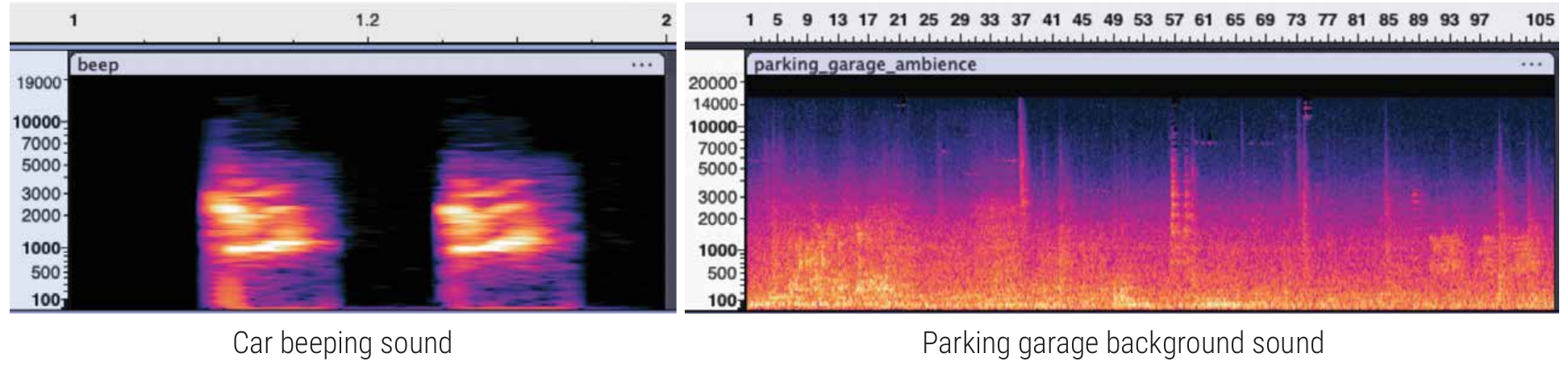}
    \captionof{figure}{Spectrogram of the car beeping sound and parking garage background noise in the data collection study.}
    \label{fig:audio-cue-spectrogram}
    \Description{Spectrogram of the car beeping sound and parking garage background noise in the data collection study.}

\end{table*}

\section{Sounds in Data Collection}
The spectrograms of the car beeping sound and background sound in the data collection study are illustrated in Figure~\ref{fig:audio-cue-spectrogram}.

\end{document}